\documentstyle[11pt]{article}
\begin{document}

\title{Interacting polytropic gas model of phantom dark energy in non-flat universe}
\author{K. Karami$^{1,2}$,\thanks{E-mail: KKarami@uok.ac.ir}\\
S. Ghaffari${^1}$\\J. Fehri${^1}$\\$^{1}$\small{Department of
Physics, University of Kurdistan, Pasdaran St., Sanandaj,
Iran}\\$^{2}$\small{Research Institute for Astronomy
$\&$ Astrophysics of Maragha (RIAAM), Maragha, Iran}\\
}

\maketitle

\begin{abstract}
By introducing the polytropic gas model of interacting dark
energy, we obtain the equation of state for the polytropic gas
energy density in a non-flat universe. We show that for even
polytropic index by choosing $K>Ba^{\frac{3}{n}}$, one can obtain
$\omega^{\rm eff}_{\Lambda}<-1$, which corresponds to a universe
dominated by phantom dark energy.
\end{abstract}
\clearpage
\section{Introduction}
Type Ia supernovae observational data deduced by two research
groups, the High-$z$ Supernovae Search Team \cite{Riess} and
 the Supernovae Cosmology Project \cite{Perlmutter}, show that the universe appears to be accelerating at
 present. Observational evidences also clear that the universe is nearly spatially flat and consists of about
 70$\%$ dark energy (DE), 30$\%$ dust matter (cold dark matter plus baryons), and negligible radiation
 \cite{Seljak}. However the nature of DE is still unknown, people
 have proposed some candidates to describe it. The cosmological
 constant, $\Lambda$, is the most obvious theoretical candidate of
 DE which has the equation of state $\omega=-1$.
 Astronomical observations indicate that the cosmological constant is many orders of magnitude
 smaller than estimated in modern theories of elementary particles \cite{Weinberg}. Also the
 "fine-tuning" and the "cosmic coincidence" problems are the two
 well known difficulties of the cosmological constant problems
 \cite{Copeland}.

 There are different alternative theories for the dynamical DE scenario which have been
 proposed by people to interpret the accelerating universe. i) The scalar-field models of DE including quintessence \cite{Wetterich,Ratra}, phantom (ghost) field \cite{Caldwell,Nojiri1,Nojiri2}, K-essence \cite{Chiba,Picon1,Picon2}
 based on earlier work of K-inflation \cite{Picon3,Garriga},
 tachyon field \cite{Sen,Padmanabhan1,Padmanabhan2}, dilatonic ghost condensate \cite{Gasperini,Hamed,Piazza},
 quintom \cite{Elizalde,Nojiri3,Anisimov}, and so forth. ii) The interacting DE models including Chaplygin gas \cite{Kamenshchik,Bento}, holographic DE models
 \cite{Cohen,Horava,Thomas,Li}, and braneworld models
 \cite{Deffayet,Sahni}, etc. The interaction between
DE and dark matter has been discussed in ample details by
\cite{Amendola1, Amendola2, Szydlowski, Tsujikawa}. The recent
evidence provided by the Abell Cluster A586 support the
interaction between dark energy and dark matter \cite{Bertolami}.
However, there are no strong observational bounds on the strength
of this interaction \cite{Feng}.

Here we study the interaction between the DE and the cold dark
matter (CDM). To do this we use the method introduced by
\cite{Setare}. Using the polytropic gas model of DE, we obtain the
equation of state for an interacting polytropic gas energy density
in a non-flat universe. We show that the interacting polytropic
gas DE in a non-flat universe behaves as a phantom type DE, i.e.
the equation of state of DE crosses the cosmological constant
boundary $\omega=-1$ during the evolution.

\section{Interacting polytropic gas}
The polytropic gas equation of state {\bf(EoS)} is given by
\begin{equation}
p_{\Lambda}=K\rho_{\Lambda}^{1+\frac{1}{n}},\label{pol1}
\end{equation}
where $K$ is a positive constant and $n$ is the polytropic index
\cite{Dalsgaard}. The different phenomenological models with EoS
as $p=-\rho+f(\rho)$ and specifically containing a polytropic part
$f(\rho)=A\rho^{\alpha}$ with constant $A$ and $\alpha$ have been
already studied in the three approaches including EoS,
scalar-tensor theory and F(R) gravity descriptions
\cite{Nojiri3,Nojiri4,Capoziello}. The polytropic EoS plays a very
important role in astrophysics. It is still very useful as simple
example which is nevertheless not too dissimilar from realistic
models. More importantly, there are cases where a polytropic EoS
is a good approximation to reality. An example is a gas where the
pressure is dominated by degenerate electrons in white dwarfs or
degenerate neutrons in neutron stars. Another example is the case
where pressure and density are related adiabatically in main
sequence stars. These motivate us to consider only the polytropic
part of $p=-\rho+f(\rho)$ for the EoS fluid description of DE in
cosmology.

Using Eq. (\ref{pol1}), the continuity equation can be integrated
to give
\begin{equation}
\rho_{\Lambda}=\left(\frac{1}{Ba^\frac{3}{n}-K}\right)^{n},\label{pol2}
\end{equation}
where $B$ is a positive integration constant and $a$ is the cosmic
scale factor. We consider a universe containing an interacting
polytropic gas energy density $\rho_{\Lambda}$ and the CDM, with
$\omega_{\rm m}=0$. The energy equations for DE and CDM are
\begin{equation}
\dot{\rho}_{\Lambda}+3H(1+\omega_{\Lambda})\rho_{\Lambda}=-Q,\label{eqpol}
\end{equation}
\begin{equation}
\dot{\rho}_{\rm m}+3H\rho_{\rm m}=Q,\label{eqCDM}
\end{equation}
where following \cite{Setare}, $Q=\Gamma\rho_{\Lambda}$ is an
interaction term and
$\Gamma=3b^2H(\frac{1+\Omega_{k}}{\Omega_{\Lambda}})$ is the decay
rate of the polytropic gas component into CDM with a coupling
constant $b^2$. Also
$\Omega_{\Lambda}=\frac{\rho_{\Lambda}}{3M^2_{\rm p}H^2}$ and
$\Omega_{k}=\frac{k}{a^2H^2}$. The choice of the interaction
between both components was to get a scaling solution to the
coincidence problem such that the universe approaches a stationary
stage in which the ratio of DE and DM becomes a constant
\cite{Hu}.

Note that choosing the $H$ in the $Q$-term is motivated purely by
mathematical simplicity. Because the continuity equations imply
that the interaction term should be a function of a quantity with
units of inverse of time (a first and natural choice can be the
Hubble factor $H$) multiplied with the energy density. Therefore,
the interaction term could be in any of the following forms: (i)
$Q\propto H\rho_{\Lambda}$ \cite{Pavon}, (ii) $Q\propto H\rho_{\rm
m}$ \cite{Guo}, or (iii) $Q\propto H(\rho_{\Lambda}+\rho_{\rm m})$
\cite{Wang1}. The dynamics of interacting DE models with different
$Q$-classes have been studied in ample details by \cite{Cabral}.
The freedom of choosing the specific form of the interaction term
$Q$ stems from our incognizance of the origin and nature of DE as
well as DM. Moreover, a microphysical model describing the
interaction between the dark components of the universe is not
available nowadays \cite{Setare2}.

Following \cite{Setare}, if we define
\begin{equation}
\omega^{\rm
eff}_{\Lambda}=\omega_{\Lambda}+\frac{\Gamma}{3H},~~~~~~\omega^{\rm
eff}_{\rm m}=-\frac{\rho_{\Lambda}}{\rho_{\rm
m}}\frac{\Gamma}{3H},
\end{equation}
then rewriting Eqs. (\ref{eqpol}) and (\ref{eqCDM}) in their
standard form reduces to
\begin{equation}
\dot{\rho}_{\Lambda}+3H(1+\omega^{\rm
eff}_{\Lambda})\rho_{\Lambda}=0,\label{seqpol}
\end{equation}
\begin{equation}
\dot{\rho}_{\rm m}+3H(1+\omega^{\rm eff}_{\rm m})\rho_{\rm
m}=0.\label{seqCDM}
\end{equation}
Following \cite{Setare}, we consider Friedmann-Robertson-Walker
(FRW) metric for the non-flat universe as
\begin{equation}
{\rm d}s^2=-{\rm d}t^2+a^2(t)\left(\frac{{\rm
d}r^2}{1-kr^2}+r^2{\rm d}\Omega^2\right),
\end{equation}
where $k$ denotes the curvature of space $k=0,1,-1$ for a flat,
closed an open universe, respectively.

Taking the time derivative of Eq. (\ref{pol2}) yields to
\begin{eqnarray}
\dot{\rho}_{\Lambda}=-3BHa^{\frac{3}{n}}\rho_{\Lambda}^{1+\frac{1}{n}}.\label{tpol2}
\end{eqnarray}
Substituting Eq. (\ref{tpol2}) in (\ref{seqpol}) gives
\begin{eqnarray}
\omega^{\rm
eff}_{\Lambda}&=&Ba^{\frac{3}{n}}\Big(3M_{\rm p}^2H^2\Omega_{\Lambda}\Big)^{\frac{1}{n}}-1,\nonumber\\
&=&-\frac{Ba^{\frac{3}{n}}}{K-Ba^{\frac{3}{n}}}-1.
\end{eqnarray}
We see that for $K>Ba^{\frac{3}{n}}$, $\omega^{\rm
eff}_{\Lambda}<-1$, which corresponds to a universe dominated by
phantom dark energy. Note that to have $\rho_{\Lambda}>0$, from
Eq. (\ref{pol2}) the polytropic index should be even,
$n=(2,4,6,\cdot\cdot\cdot)$. Also note that for $a\rightarrow
a_s=(\frac{K}{B})^{\frac{n}{3}}$ then Eqs.
(\ref{pol1})-(\ref{pol2}) show that
$\rho_{\Lambda}\rightarrow\infty$ and
$p_{\Lambda}\rightarrow\infty$. Therefore the selected polytropic
DE model shows a finite-time future singularity type III. The
properties of different future singularities in the DE universe
have been investigated in ample details by \cite{Nojiri3,Nojiri5}.

Following \cite{Copeland}, one can obtain a corresponding
potential for the polytropic gas by treating it as an ordinary
scalar field $\phi(t)$. Using Eqs. (\ref{pol1}), (\ref{pol2})
together with $\rho_{\phi}=\frac{1}{2}\dot{\phi}^2+V(\phi)$ and
$p_{\phi}=\frac{1}{2}\dot{\phi}^2-V(\phi)$, we find
\begin{eqnarray}
\dot{\phi}^2&=&Ba^{\frac{3}{n}}\Big(3M_{\rm p}^2H^2\Omega_{\Lambda}\Big)^{1+\frac{1}{n}},\nonumber\\
&=&-Ba^{\frac{3}{n}}\left(\frac{1}{K-Ba^\frac{3}{n}}\right)^{n+1},~~~n=(2,4,6,\cdot\cdot\cdot),\label{phidot}
\end{eqnarray}

\begin{equation}
V(\phi)=\frac{K-\frac{B}{2}~a^\frac{3}{n}}{\Big(K-Ba^\frac{3}{n}\Big)^{n+1}}~,~~~n=(2,4,6,\cdot\cdot\cdot).\label{Vphi}
\end{equation}

Equation (\ref{phidot}) shows that for $K>Ba^{\frac{3}{n}}$,
$\dot{\phi}^2<0$. Therefore by definition $\phi=i\psi$, the
Lagrangian of the scalar field $\phi(t)$ can be rewritten as
\begin{equation}
L=\frac{1}{2}\dot{\phi}^2-V(\phi)=-\frac{1}{2}\dot{\psi}^2-V(i\psi).
\end{equation}
The energy density and the pressure corresponding to the scalar
field $\psi$ are as follows, respectively:
\begin{eqnarray}
\rho_{\psi}&=&-\frac{1}{2}\dot{\psi}^2+V(i\psi),\nonumber\\
p_{\psi}&=&-\frac{1}{2}\dot{\psi}^2-V(i\psi).
\end{eqnarray}
One can conclude that the scalar field $\psi$ is a phantom field.
Therefore, a phantom-like equation of state can be generated from
an interacting polytropic gas DE model in a non-flat universe.

To obtain a corresponding potential for the polytropic gas we
start from the first Friedmann equation given by
\begin{equation}
H^2=\frac{1}{3M^2_{\rm p}}(\rho_{\rm
m}+\rho_{\Lambda})-\frac{k}{a^2}.\label{F1}
\end{equation}
Since $p_{\rm m}=0$ for the CDM, hence the continuity equation can
be integrated to give
\begin{equation}
\rho_{\rm m}=\frac{\rho_{\rm m_0}}{a^3}\label{rhom}.
\end{equation}
Substituting Eqs. (\ref{pol2}) and (\ref{rhom}) in Eq. (\ref{F1})
yields to
\begin{equation}
H=\left\{\frac{1}{3M^2_{\rm p}}\Big[\frac{\rho_{\rm
m_0}}{a^3}+\Big(K-Ba^\frac{3}{n}\Big)^{-n}\Big]-\frac{k}{a^2}\right\}^{1/2}.\label{F2}
\end{equation}
Using Eq. (\ref{F2}), definition $\phi=i\psi$ and
$\frac{d\psi}{da}=\frac{\dot{\psi}}{aH}$, we can rewrite Eq.
(\ref{phidot}) in terms of the integral of $a$ as
\begin{equation}
\psi=-\frac{2nM_{\rm p}}{\sqrt{3}}\int\Big[\rho_{\rm
m_0}B^n(K-u^2)^{1-n}u^{2n}-3kM^2_{\rm
p}B^{2n/3}(K-u^2)^{1-\frac{2n}{3}}u^{2n}+(K-u^2)\Big]^{-1/2}du,\label{psi}
\end{equation}
where $u:=\sqrt{K-Ba^\frac{3}{n}}$.

Now if we consider a flat universe containing only the polytropic
gas, i.e. $\rho_{\rm m_0}\rightarrow 0$ and $k\rightarrow 0$, then
Eq. (\ref{psi}) can be easily integrated to give
\begin{eqnarray}
\psi&=&-\frac{2nM_{\rm
p}}{\sqrt{3}}\sin^{-1}\Big(\frac{u}{\sqrt{K}}\Big),\nonumber\\
&=&-\frac{2nM_{\rm
p}}{\sqrt{3}}\sin^{-1}\Big(\sqrt{1-Ba^\frac{3}{n}/K}~\Big),
\end{eqnarray}
or
\begin{equation}
a^\frac{3}{n}=\frac{K}{B}\cos^2\Big(\frac{\sqrt{3}}{2nM_{\rm
p}}\psi\Big).
\end{equation}
Substituting this for Eq. (\ref{Vphi}), we obtain the following
potential
\begin{equation}
V(i\psi)=\frac{1}{2K^n}\frac{1+\sin^2\Big(\frac{\sqrt{3}}{2nM_{\rm
p}}\psi\Big)}{\Big[\sin^{2}\Big(\frac{\sqrt{3}}{2nM_{\rm
p}}\psi\Big)\Big]^{n+1}}~,~~~n=(2,4,6,\cdot\cdot\cdot).
\end{equation}
Hence, a coupled filed with this potential is equivalent to the
polytropic gas model.
\section{Conclusions}
Here we have introduced a polytropic gas model of DE as an
alternative model to explain the accelerated expansion of the
universe. We have considered a non-flat FRW universe filled with
this fluid which is in interact with the CDM. We showed that the
interacting polytropic gas DE with even polytropic index and
$K>Ba^{\frac{3}{n}}$ can be described by a phantom field which has
$\omega^{\rm eff}_{\Lambda}<-1$. Note that in the present work
regarding the selected polytropic EoS, there are two things that
have been investigated in ample details in comparison with other
works in the literature. i) The interaction between the polytropic
DE and the CDM which can not only solve the coincidence problem
but also cause the EoS of DE crosses the cosmological constant
boundary $\omega=-1$ during the evolution. ii) The non-flatness of
the polytropic universe. However some experimental data have
implied that our universe is not a perfectly flat universe and it
possess a small positive curvature ($\Omega_k\sim 0.015$)
\cite{Spergel}. Although it is believed that our universe is flat,
a contribution to the Friedmann equation from spatial curvature is
still possible if the number of e-foldings is not very large
\cite{Setare3}.
\\
\\
\noindent{{\bf Acknowledgements}}. This work has been supported
financially by Research Institute for Astronomy $\&$ Astrophysics
of Maragha (RIAAM), Maragha, Iran.


\end{document}